\documentclass[twoside,twocolumn,english,prc,amsmath,a4paper,myheadings]{revtex4}
\usepackage[T1]{fontenc}
\usepackage{geometry}
\usepackage{amsmath}
\usepackage{graphicx}
\usepackage{amssymb}

\makeatletter
\def\@oddhead{\rightmark \hfill Evidence for flow in pPb collisions  \hfill \thepage}
\def\@evenhead{\thepage \hfill K. Werner et al.\hfill}
\def\fnum@table{\tablename~{\bf\thetable}}
\def\fnum@figure{\figurename~{\bf\thefigure}}
\def\tablename{\footnotesize{\bf Table}}
\def\figurename{\footnotesize{\bf Figure}}

\usepackage{dcolumn}

\def\citet{\cite}
\newcommand{\dd}{\partial}

\makeatother

\usepackage{babel}

\begin{document}

\title{Evidence for flow in pPb collisions at 5 TeV from v$_{\mathbf{2}}$
mass splitting }

\author{{\normalsize K.$\,$Werner$^{(a)}$, M. Bleicher$^{(b)}$, B. Guiot$^{(a)}$,
Iu.$\,$Karpenko$^{(b,c)}$, T.$\,$Pierog$^{(d)}$ }}

\address{$^{(a)}$ SUBATECH, University of Nantes -- IN2P3/CNRS-- EMN, Nantes,
France}

\address{$^{(b)}$ FIAS, Johann Wolfgang Goethe Universitaet, Frankfurt am
Main, Germany}

\address{$^{(c)}$ Bogolyubov Institute for Theoretical Physics, Kiev 143,
03680, Ukraine}

\address{$^{(d)}$Karlsruhe Inst. of Technology, KIT, Campus North, Inst.
f. Kernphysik, Germany}

\begin{abstract}
We show that a fluid dynamical scenario describes quantitatively the
observed mass splitting of the elliptical flow coefficients $v_{2}$
for pions, kaons, and protons. This provides a strong argument in
favor of the existence of a fluid dynamical expansion in pPb collisions
at 5TeV.
\end{abstract}
\maketitle
One of the strongest signals of collective flow in heavy ion collisions
is the fact that the transverse momentum dependence of the elliptical
flow coefficient $v_{2}$ (measuring the azimuthal asymmetry) depends
in a very characteristic way on the mass of the observed hadrons.
This has been predicted \citet{pasi} and impressively confirmed experimentally
later \citet{intro1,intro2}. Can we also {}``prove'' the existence
of flow in small systems like proton-lead collisions, where such a
collective behavior has not been expected?

Information about flow asymmetries can be obtained via studying two
particle correlations as a function of the pseudorapidity difference
$\Delta\eta$ and the azimuthal angle difference $\Delta\phi$. So-called
ridge structures (very broad in $\Delta\eta$) have been observed
first in heavy ion collisions, later also in pp \citet{cms_ridge}
and very recently in pPb collisions \citet{alice1,cms,atlas}. In
case of heavy ions, these structures appear naturally in models employing
a hydrodynamic expansion, in an event-by-event treatment. 

To clearly pin down the origin of such structures in small systems,
one needs to consider identified particles. In the fluid dynamical
scenario, where particles are produced in the local rest frame of
fluid cells characterized by transverse velocities, large mass particles
(compared to low mass ones) are pushed to higher transverse momenta,
visible in $p_{t}$ distributions, but also in the dihadron correlations.
Both effects are clearly seen in experimental data. In this paper,
we focus on dihadron correlations and the $v_{2}$ coefficients.

The following discussion is based on ALICE results on dihadron correlations
\citet{alice1,alice2} and EPOS3 simulations. EPOS3 is a major update
of the work described in \citet{jetbulk} (EPOS2). Here, we introduced
a theoretical scheme which accounts for hydrodynamically expanding
bulk matter, jets, and the interaction between the two. The whole
transverse momentum range is covered, from very low to very high $p_{t}$.
In \citet{jetbulk}, we show that this approach can accommodate spectra
of jets with $p_{t}$ up to 200 GeV/c in $pp$ scattering at 7 TeV,
as well as particle yields and harmonic flows with $p_{t}$ between
0 and 20 GeV/c in PbPb collisions at 2.76 TeV. To our knowledge, this
is the only model able to describe the famous lambda to kaon enhancement
correctly over the whole $p_{t}$ range. New features of EPOS3 are:
an event-by-event 3D+1 viscous hydrodynamical evolution, and a new
treatment of high parton densities, via an individual parton saturation
scale $Q_{s}$ for each elementary scattering. More details will be
given in the appendix. All results in this paper are based on EPOS3.074.

\begin{figure}[b]
\begin{centering}
\includegraphics[angle=270,scale=0.32]{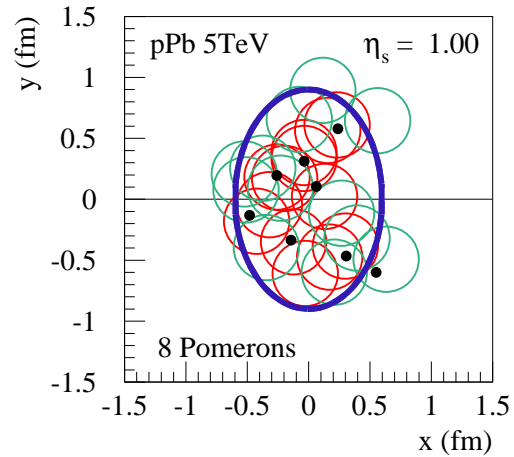}
\par\end{centering}

\caption{(Color online) Example of a semi-peripheral p-Pb scattering, with
8 Pomerons, showing the transverse plane at space-time rapidity $\eta_{s}=1$.
The positions of the Pomerons are indicated by the black dots. String
segment having enough energy to escape (corona) are marked green,
the red ones constitute the core.\label{fig:eibp2}}

\end{figure}

As in \citet{jetbulk}, the basis of our approach is multiple scattering
(even for $pp$), where a single scattering is a hard elementary scattering
plus initial state radiation, the whole object being referred to as
parton ladder or Pomeron. The corresponding final state partonic system
amounts to (usually two) color flux tubes, being mainly longitudinal,
with transversely moving pieces carrying the $p_{t}$ of the partons
from hard scatterings. These flux tubes constitute eventually both
bulk matter, also referred to as {}``core'' (which thermalizes,
flows, and finally hadronizes) and jets (also refereed to as {}``corona''),
according to some criteria based on the energy of the string segments
and the local string density. 

In p-Pb collisions, the {}``violence'' of a collision is no longer
characterized by the impact parameter, but by the multiplicity, or
in the multiple scattering approach, by the number of Pomerons --
being proportional to the multiplicity. In fig. \ref{fig:eibp2},
we show a simulation of a semi-peripheral p-Pb scattering, with 8
Pomerons. We consider the transverse plane at space-time rapidity
$\eta_{s}=1$. The positions of the Pomerons are indicated by the
black dots. String segment having enough energy to escape are marked
green, the red ones constitute the core.

\begin{figure}[t]
\begin{centering}
\includegraphics[scale=0.25]{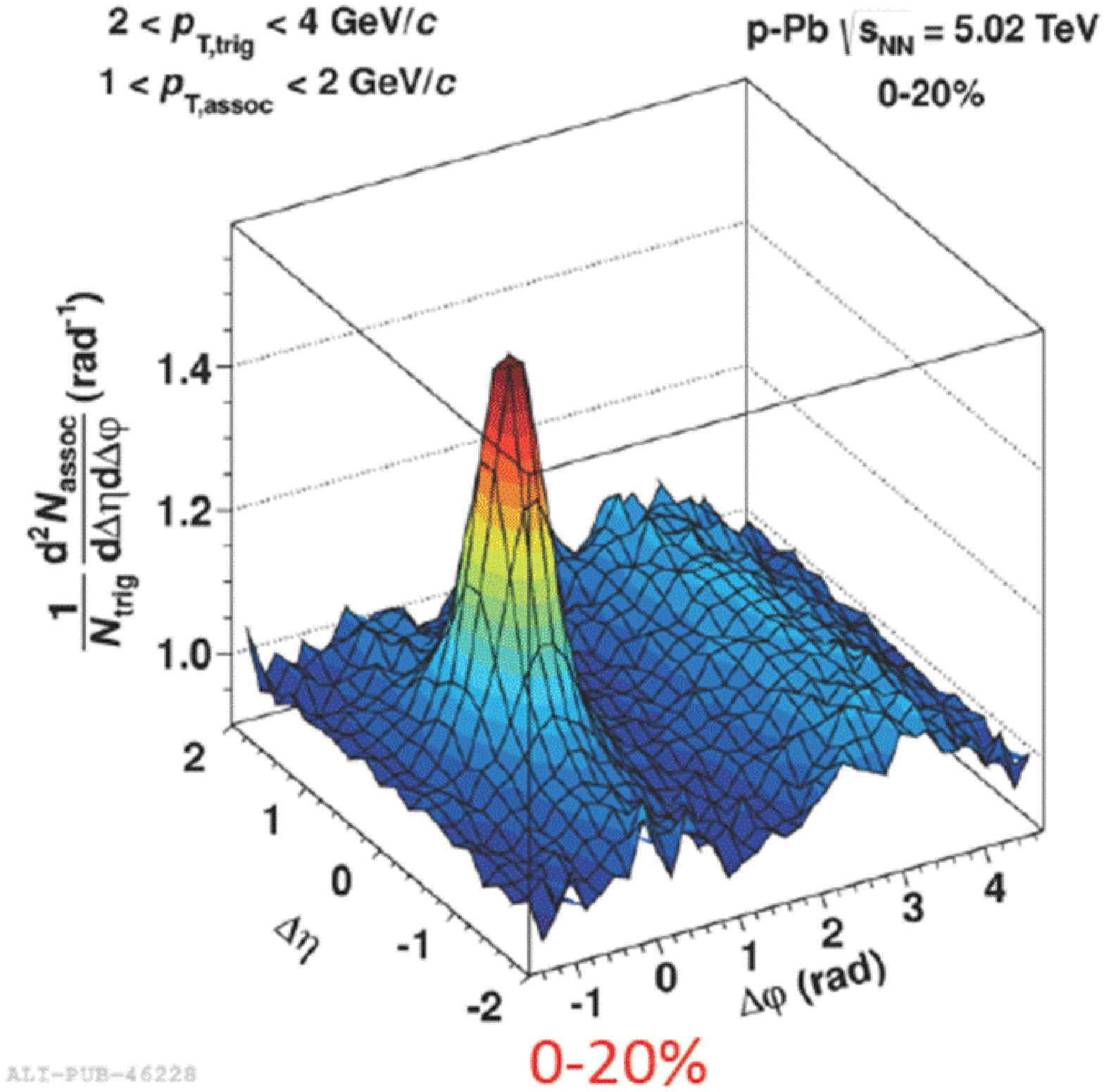}\\
\includegraphics[scale=0.32]{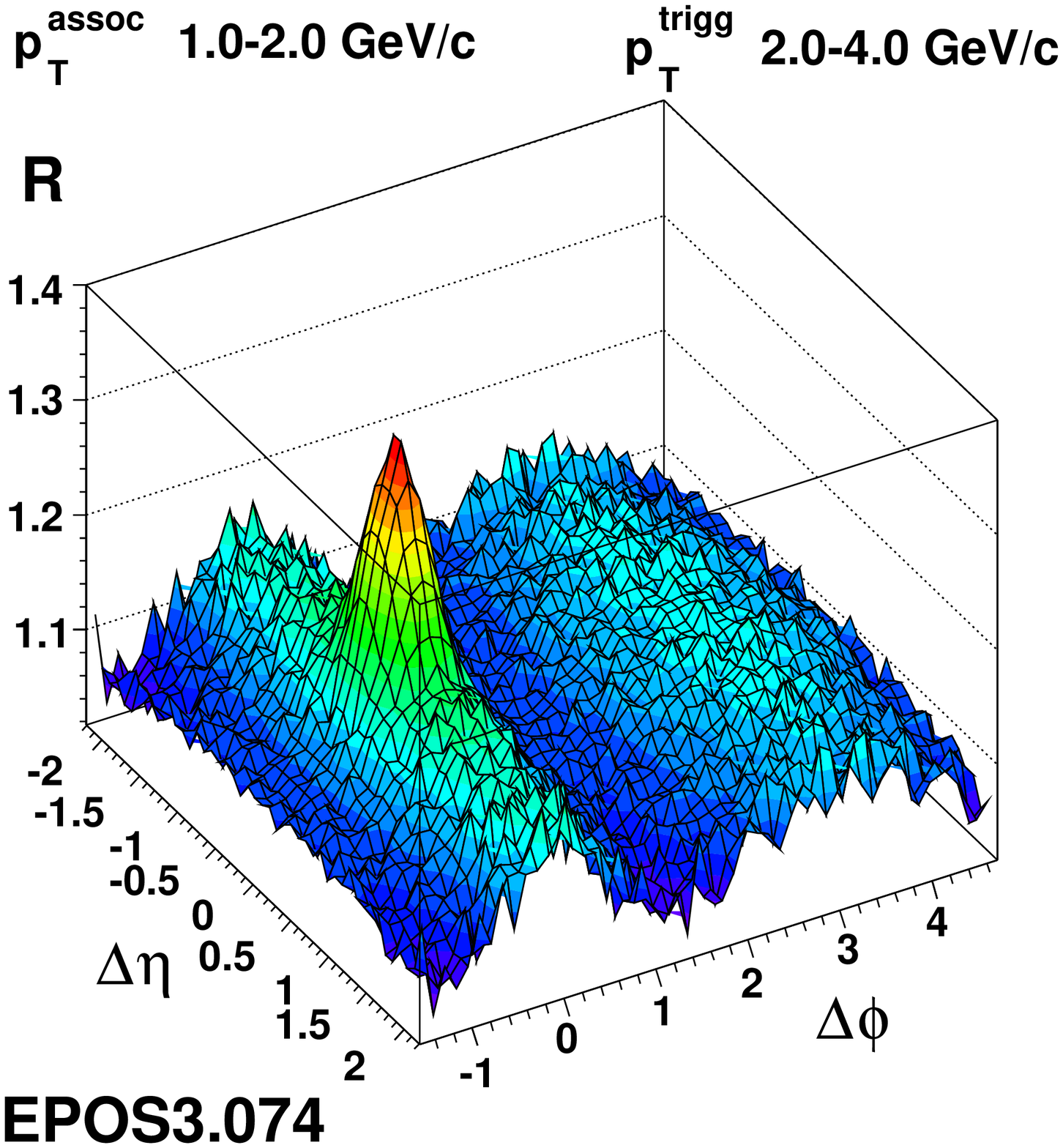}
\par\end{centering}

\caption{(Color online) Dihadron correlation function for charged particles,
from ALICE (upper plot) and EPOS3 simulations, for high multiplicity
(0-20\%) events.\label{fig:dih1}}

\end{figure}

The positions of the Pomerons, and therefore also the positions of
the string segments, are generated randomly. In the current example,
the overall shape of the core part (the red circles) happens to be
elongated along the $y$ axis, so we have generated randomly an elliptical
shape (indicated by the blue ellipse, to guide the eye). This will
lead to a preferred expansion along the $x$ axis, and a $\cos(2\Delta\phi)$
shape of the correlation function. The question is only, if the fluctuations
are sufficiently big and sufficiently frequent to provide a quantitative
agreement with the observed correlation functions. In fig. \ref{fig:dih1},
we plot the dihadron correlation function for charged particles, from
ALICE (upper plot) and EPOS3 simulations. We observe in theory and
experiment the {}``jet peak'' as well as a near side and a awayside
ridge structure. 

\begin{figure}[t]
\begin{centering}
\includegraphics[scale=0.25]{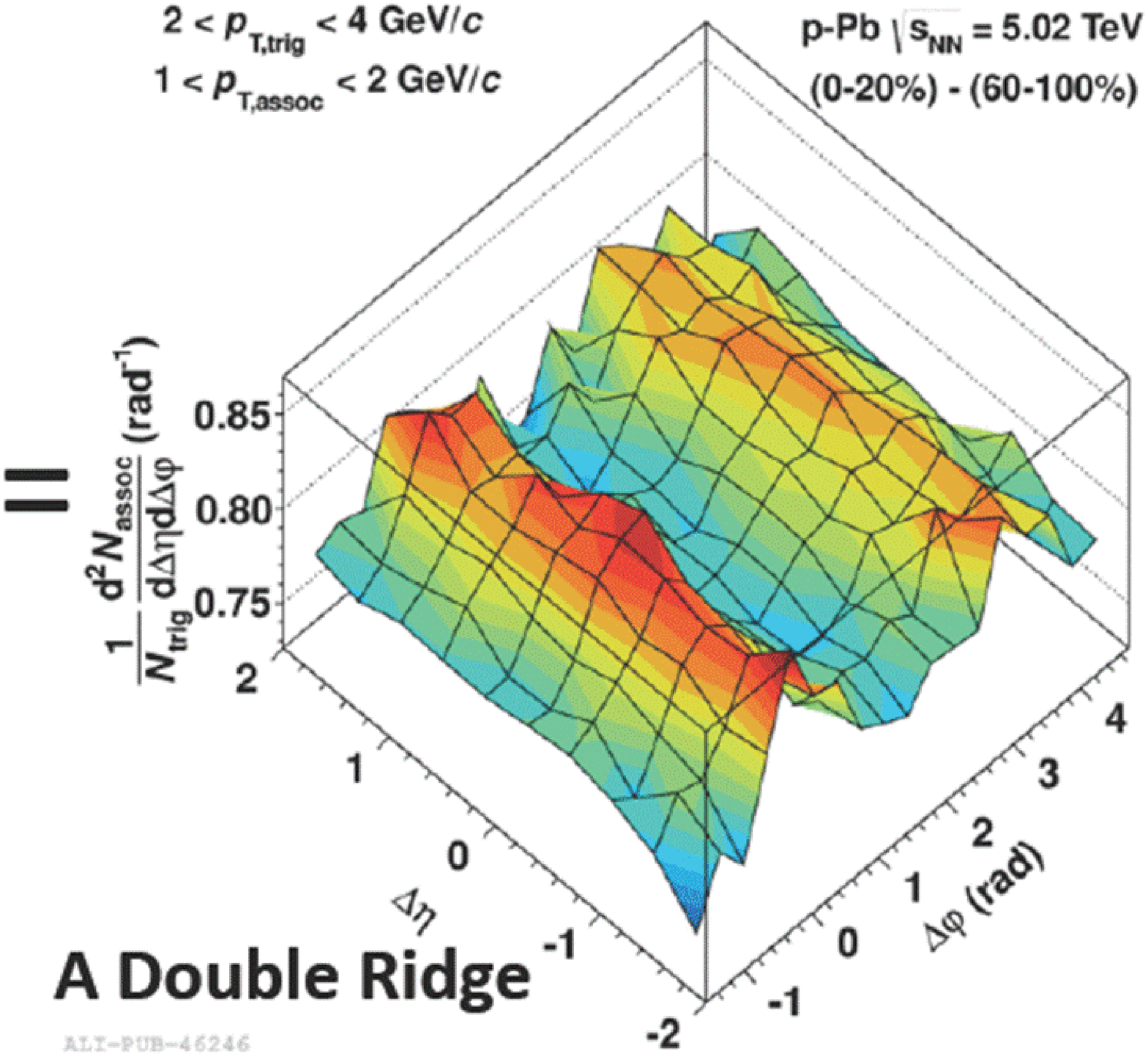}\\
\includegraphics[scale=0.32]{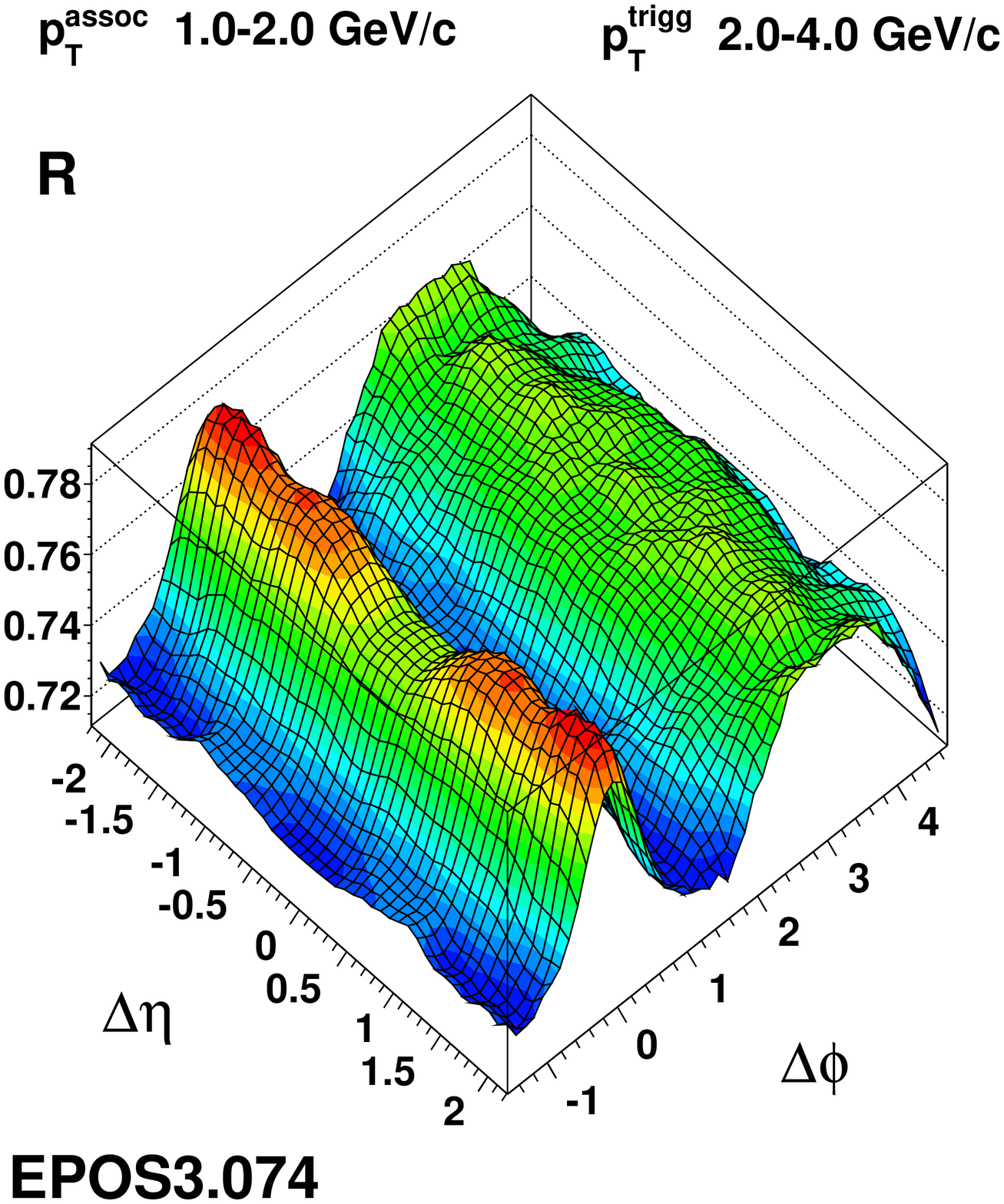}
\par\end{centering}

\caption{(Color online) Dihadron correlation function for charged particles,
from ALICE (upper plot) and EPOS3 simulations, for high multiplicity
(0-20\%) events, after subtraction of the result for the 60-80\% class.\label{fig:dih2}}

\end{figure}

To get rid of, the jet correlations, one subtracts the low multiplicity
correlation function (assuming that the jet correlation is the same).
In fig. \ref{fig:dih2}, %
we show the results for ALICE measurement and the corresponding EPOS3
simulations. Indeed, a double ridge structure is visible, essentially
flat in the $\Delta\eta$ directions. The origin of the double ridge
are random azimuthal asymmetries, as shown in fig. \ref{fig:eibp2},
which lead to asymmetric flow and to a double peak structure in $\Delta\phi$.
The translational invariance of the structure finally leads to a double
ridge. The simulation result is an absolute prediction, it is the
yield of associated particles per trigger (and per $\Delta\eta\Delta\phi$.
So indeed, the magnitude of the effect is of the right order.

\begin{figure}[tb]
\begin{centering}
\includegraphics[angle=270,scale=0.32]{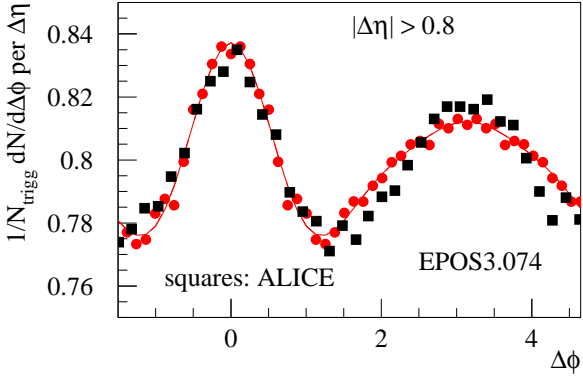}
\par\end{centering}

\caption{(Color online) Associated yield per trigger, projected onto $\Delta\phi$,
for $|\Delta\eta|>0.8$. We show ALICE results (black squares) and
EPOS3 simulations (red dots). \label{fig:proj}}

\end{figure}

For a more quantitative analysis, one considers the projections onto
$\Delta\phi$, for $|\Delta\eta|>0.8$, as shown in fig. \ref{fig:proj}
for ALICE measurement and the corresponding EPOS3 simulations (the
latter ones have been multiplied by 1.07). The solid red line represents
a Fourier decomposition,\begin{equation}
\frac{1}{N_{\mathrm{trig}}}\,\frac{dN}{d\Delta\phi}\{per\,\Delta\eta\}=\sum_{n=0}^{5}2a_{n}\cos(n\Delta\phi).\label{eq:modu}\end{equation}

\begin{figure}[tb]
\begin{centering}
\includegraphics[angle=270,scale=0.32]{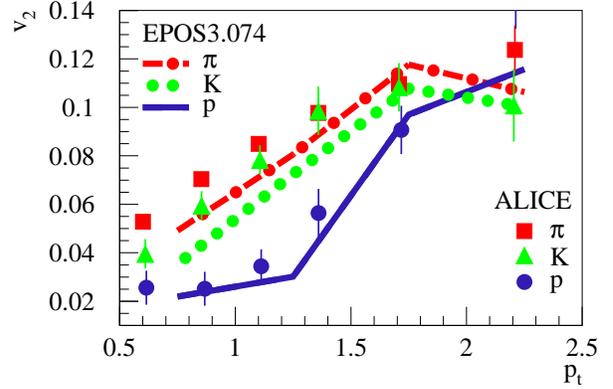}
\par\end{centering}

\caption{(Color online) Elliptical flow coefficients $v_{2}$ for pions, kaons,
and protons. We show ALICE results (squares) and EPOS3 simulations
(lines). Pions appear red, kaons green, protons blue. \label{fig:v2}}

\end{figure}

The same procedure can be done for identified particles. We consider
four types of pairs: $h-h$, $h-\pi$, $h-K$, and $h-p$, where $h$
stands for charged hadrons, $\pi$ for charged pions, $K$ for charged
kaons, and $p$ for protons and antiprotons. For each of the four
cases one computes the Fourier coefficients $a_{n}$. To normalize
the coefficients, we have to divide by the so-called {}``baseline''
$b$. We follow precisely the ALICE prescription: the baseline is
the average yield of the high multiplicity correlation function, extracted
from the latter one at $\pi/2$, taking into account the modulation
of the functions following eq. (\ref{eq:modu}). In this way we obtain
the normalized Fourier coefficients $v_{n\Delta}$, for the four types
of pairs considered, i.e. $v_{n\Delta}^{h-h}$, $v_{n\Delta}^{h-\pi}$,
$v_{n\Delta}^{h-K}$, $v_{n\Delta}^{h-p}$. The flow harmonics, characterizing
inclusive identified particle production from a {}``flowing'' medium,
are then given as\begin{equation}
v_{n}^{\pi}=\frac{v_{n\Delta}^{h-\pi}}{\sqrt{v_{n\Delta}^{h-h}}},\; v_{n}^{K}=\frac{v_{n\Delta}^{h-K}}{\sqrt{v_{n\Delta}^{h-h}}},\; v_{n}^{p}=\frac{v_{n\Delta}^{h-p}}{\sqrt{v_{n\Delta}^{h-h}}}.\;\end{equation}
The results for the second harmonics (elliptical flow) are shown in
fig. \ref{fig:v2}. %
\begin{figure}[b]
\begin{centering}
\includegraphics[angle=270,scale=0.32]{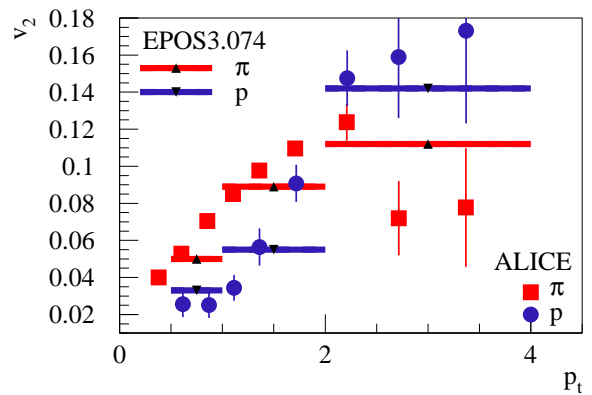}
\par\end{centering}

\caption{(Color online) Elliptical flow coefficients $v_{2}$ for pions and
protons. We show ALICE results (squares) and EPOS3 simulations (horizontal
lines). Pions appear red, protons blue. \label{fig:v2b}}

\end{figure}
Clearly visible in data and in the simulations: a separation of the
results for the three hadron species: in the $p_{t}$ range of 1-1.5
GeV/c, the kaon $v_{2}$ is somewhat below the pion one, whereas the
proton result is clearly below the two others. To discuss higher values
of $p_{t}$ (and due to limited statistics), we use a different binning
in $p_{t}$ (0.5-1, 1-2, 2-4 GeV/c), see fig. \ref{fig:v2b}. We compare
pion and proton results, and we clearly see (in data and simulations)
a {}``crossing'': the $v_{2}$ of protons is below the one of pions,
below $2$GeV/c, and above beyond $2$GeV/c.

Within our fluid dynamical approach, the above results are nothing
but a {}``mass splitting''. The effect is based on an asymmetric
(mainly elliptical) flow, which translates into the corresponding
azimuthal asymmetry for particle spectra. Since a given velocity translates
into momentum as $m_{A}\gamma v$, with $m_{A}$ being the mass of
hadron type $A$, flow effects show up at higher values of $p_{t}$
for higher mass particles.

To summarize: We have shown that a realistic fluid dynamical scenario
describes quantitatively the observed mass splitting of the elliptical
flow coefficients $v_{2}$ for pions, kaons, and protons. This provides
a strong argument in favor of the existence of a fluid dynamical expansion
in proton-lead collisions at 5TeV.

\appendix

\section{Gribov Regge approach with saturation scales}

Gribov-Regge approach here means an assumption about the structure
of the $T$ matrix, expressed in terms of elementary objects called
Pomerons, occurring in parallel. Squaring the matrix element, the
total cross section can be expressed as 

\begin{minipage}[c][1\totalheight]{0.4\columnwidth}%
\noindent \begin{center}
\[
\sigma^{\mathrm{tot}}=\sum_{\mathrm{cut\, P}}\int\sum_{\mathrm{uncut\, P}}\int\]

\par\end{center}%
\end{minipage}%
\begin{minipage}[c][1\totalheight]{0.5\columnwidth}%
\noindent \begin{center}
\includegraphics[scale=0.2]{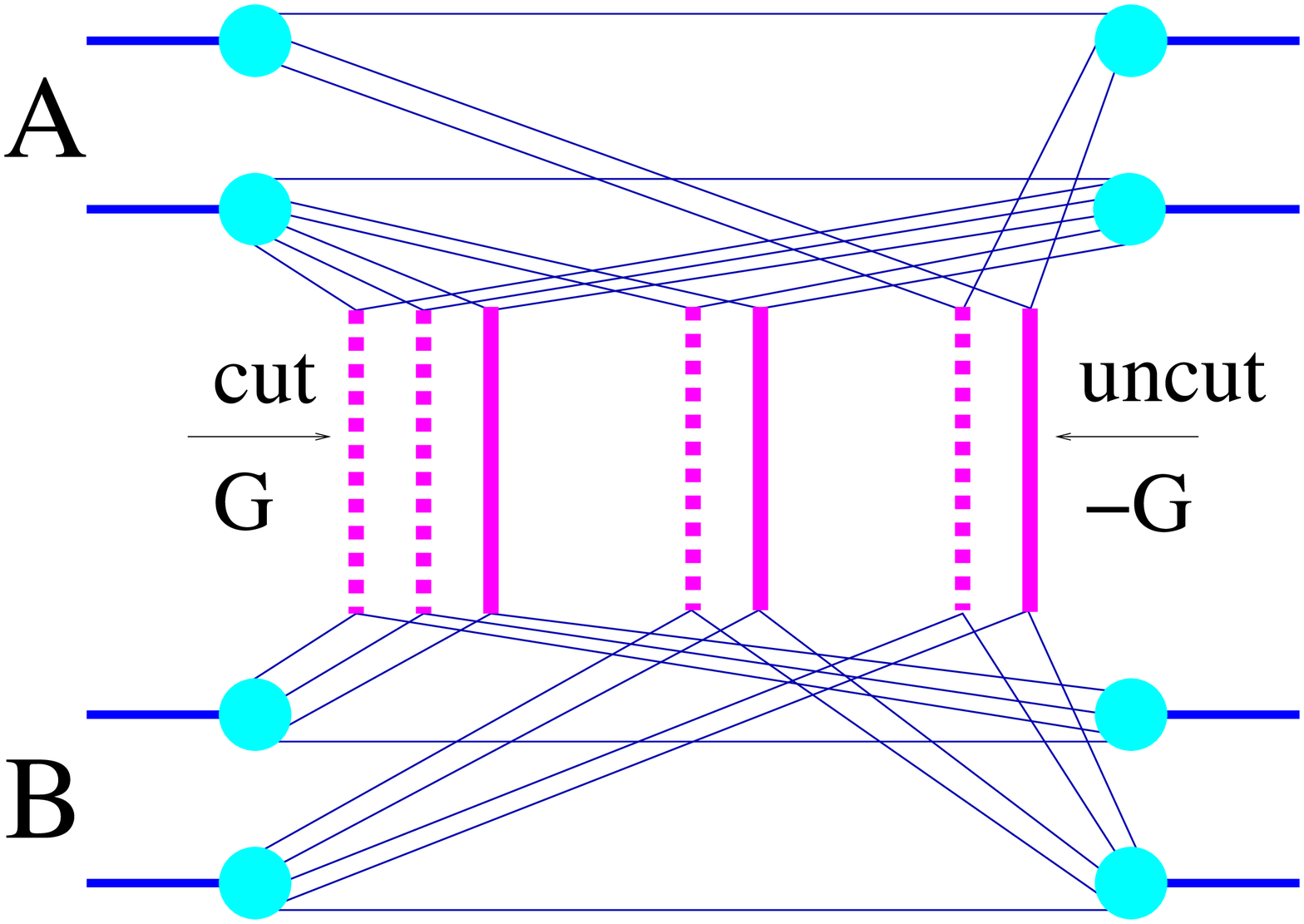}
\par\end{center}%
\end{minipage}\\
in terms of cut and uncut Pomerons, for pp, pA, and AA. Partial summation
provides exclusive cross sections. The diagram corresponds to precisely
defined mathematical expressions, see \citet{hajo}. The expression
for a cut Pomeron is \[
G=\frac{1}{2\hat{s}}2\mathrm{Im}\,\{{\cal FT}\{T\}\}(\hat{s},b),\: T=i\hat{s}\,\sigma_{hard}(\hat{s})\,\exp(R_{\mathrm{hard}}^{2}t),\]
where $FT$ means Fourier transform. The parton-parton cross section
$\sigma_{\mathrm{hard}}$ is computed based on DGLAP parton evolutions
from both sides and a Born pQCD hard scattering, restricting the virtualities
to values bigger than a saturation scale $Q_{s}\propto N_{\mathrm{part}}\hat{s}^{\lambda}$.
This scale is individual for each Pomeron, $N_{\mathrm{part}}$ is
the number of participating nucleons attached to this Pomeron, $\hat{s}$
is its energy.

\section{Viscous hydrodynamics}

In EPOS3, we employ an event-by-event treatment of a numerical solution
in 3D+1 dimensions of the hydrodynamic equations (in Israel-Stewart
formulation), using Milne coordinates \citet{yuri},\begin{eqnarray*}
\dd_{;\nu}T^{\mu\nu} & = & \dd_{\nu}T^{\mu\nu}+\Gamma_{\nu\lambda}^{\mu}T^{\nu\lambda}+\Gamma_{\nu\lambda}^{\nu}T^{\mu\lambda}=0\,,\\
\gamma\left(\dd_{t}+v_{i}\dd_{i}\right)\pi^{\mu\nu} & = & -\frac{1}{\tau_{\pi}}\left(\pi^{\mu\nu}-\pi_{\text{NS}}^{\mu\nu}\right)+I_{\pi}^{\mu\nu}\,,\\
\gamma\left(\dd_{t}+v_{i}\dd_{i}\right)\Pi & = & -\frac{1}{\tau_{\Pi}}\left(\Pi-\Pi_{\text{NS}}\right)+I_{\Pi}\,,\end{eqnarray*}
where $\pi^{\mu\nu}$ and $\Pi$ are the shear stress tensor and bulk
pressure, respectively, $\Delta^{\mu\nu}=g^{\mu\nu}-u^{\mu}u^{\nu}$
is the projector orthogonal to $u^{\mu}$, and 

$T^{\mu\nu}=\epsilon u^{\mu}u^{\nu}-(p+\Pi)\Delta^{\mu\nu}+\pi^{\mu\nu}$, 

$\pi_{\text{NS}}^{\mu\nu}=\eta(\Delta^{\mu\lambda}\dd_{;\lambda}u^{\nu}+\Delta^{\nu\lambda}\dd_{;\lambda}u^{\mu})-\frac{2}{3}\eta\Delta^{\mu\nu}\dd_{;\lambda}u^{\lambda}$,

$\Pi_{\text{NS}}=-\zeta\dd_{;\lambda}u^{\lambda}$,

$I_{\pi}^{\mu\nu}=-\frac{4}{3}\pi^{\mu\nu}\dd_{;\gamma}u^{\gamma}-[u^{\nu}\pi^{\mu\beta}+u^{\mu}\pi^{\nu\beta}]u^{\lambda}\dd_{;\lambda}u_{\beta}$,

$I_{\Pi}=-\frac{4}{3}\Pi\dd_{;\gamma}u^{\gamma}$.

\noindent We use for the calculation in this paper always $\eta/S=0.08$,
$\zeta/S=0.$

\end{document}